\documentclass[aps,pra,onecolumn,preprint,preprintnumbers,groupaddress,amsmath,amssymb,usenatbib,
nofootinbib,showkeys,showpacs,altaffilletter,11pt]{revtex4}
\usepackage[dvips]{graphicx}
\usepackage{dcolumn}
\usepackage{bm}
\usepackage{latexsym}
\usepackage{amsfonts}
\usepackage{subfigure}
\usepackage{amssymb}
\usepackage{amsmath}
\usepackage[colorlinks]{hyperref}
\usepackage{color}

\begin{document}

\title{Interacting $F(R,T)$ gravity with modified Chaplygin gas}

\author{Ali R. Amani}
\email{a.r.amani@iauamol.ac.ir}
\affiliation{\centerline{Faculty of Sciences, Department of Physics, Ayatollah Amoli Branch, Islamic Azad University,}\\ P.O. Box 678, Amol, Mazandaran, Iran}

\author{S. L. Dehneshin}
\email{ladan\_dehneshin@yahoo.com}
\affiliation{\centerline{Department of Physics, Central Tehran Branch, Islamic Azad University, Tehran, Iran.}}

\date{\today}

\begin{abstract}
In this paper, we have studied $F(R,T)$ gravity as an arbitrary function of curvature and torsion scalars in Friedmann--Lema\^{\i}tre--Robertson--Walker (FLRW) background. Then, we have considered interacting model between $F(R,T)$ gravity and modified Chaplygin gas. The novelty of this model is that the Universe includes both cases curvature and torsion, and one dominated by a Chaplygin gas. In order to calculate cosmological solutions, we obtained Friedmann equations and also equation of state (EoS) parameter of dark energy. By employing interacting model we considered the total energy density and the total pressure of Universe as the combination of components of dark energy and Chaplygin gas. Subsequently, we reconstructed the model by an origin of a scalar field entitled quintessence model with a field potential. The field potential has been calculated in terms of free parameters of  $F(R,T)$ gravity and modified Chaplygin gas. In what follows, we used a parametrization, and the cosmological parameters have been written in terms of redshift $z$. Next, we plotted cosmological parameters with respect to three variable of cosmic time, redshift $z$ and $e$-folding number $N=ln(a)$, and the figures showed us an accelerated expansion of Universe. Also, we have described the scenario in three status early time, late time and future time by $e$-folding number. Finally, the stability of scenario has been investigated by a useful function named sound speed, and the graph of sound speed versus $e$-folding number has been showed us that there is the stability in late time.
\end{abstract}
\maketitle

\section{Introduction}\label{s1}
As we know, the Universe is expanding and this expanding is undergoing an accelerating phase. This issue discovered in type Ia supernova \cite{Riess_1998, Perlmutter_1999, Bachall_1999}, associated with large scale structure  \cite{Tegmark_2004, Abazajian_2004, Pope_2004} and cosmic microwave background  \cite{Bennet_2003, Spergel_2003}. It is noteworthy that the accelerated expansion of the Universe is created from a mysterious energy called dark energy, which is about Two thirds the total energy of Universe. By using the Einstein field equation, the accelerated expansion described by a small positive cosmological constant. Also this discovery demonstrates that geometry of the Universe is very close to flat space \cite{GOMG}. Various candidates have been introduced to describe dark energy scenario, so that the Universe dominates with a perfect fluid by a negative pressure and the EoS parameter which is less than $-1$, the so-called phantom phase. We can introduce some of these models such as the cosmological constant \cite{WEIS,PEEP}, the scalar fields ( including quintessence, $K$-essence, phantom, tachyon, etc) \cite{Zlatev_1999, Kamenshchik_2001, Caldwell_2002, Amani_2011, Amani_2009, Amani1_2009}, the vector field \cite{Chiba_2008, Sadeghi_2010}, holographic \cite{Amani1_2011, Wei_2009, Farooq_2010}, interacting model \cite{Amani_2013, Amani_2014} and braneworld models \cite{Sahni_2003, Setarea_2009, Setare_2008}.

Moreover, there are some other methods to describe the Universe accelerated expansion. One of these theories is modified gravity theories that from the physical point of view, this prevents the complexities of the previous models especially the complicated computation of numerical solutions. Another benefit of the modified gravity theories is consistent with recent observations for late accelerating Universe and dark energy. Some of these modified gravity theories can directly be achieved by replacing the Ricci scalar $R$ by $F (R)$, $F(T)$ and $F(G)$ with an arbitrary function in the Einstein-Hilbert action. These theories are introduced as $F(R)$ gravity \cite{Nojiri_2006, Nojiri_2007},  $F(T)$ gravity \cite{Linder_2010, Myrzakulov_2011, Chen_2011, Dent_2011, Harko_2014} and $F(G)$ gravity \cite{Nojiri_2005, Li_2007, Kofinas_2014} which respectively an arbitrary function of the Ricci scalar $R$, the torsion scalar $T$ and the Gauss-Bonnet term $G$. Therefore, we can tell that modified gravitational theories are a generalization of general relativity. We also note that $F(T)$ gravity is a generalized version of teleparallel gravity originally proposed by Einstein \cite{Einstein_1928}, he had tried to redefine the unification of gravity and electromagnetism on the mathematical structure of distant parallelism by introducing of a tetrad or vierbein field, the so-called teleparallelism. Thus, instead of using the Levi-Civita connection in the framework of general relativity, we use the Weitzenb\"{o}ck connection in teleparallelism. In that case, the four-dimensional space-time manifold must be a parallelizable manifold \cite{Weitzenbock_1923, Bengochea_2009}.

In this paper, we will explain the late time accelerated expansion of the Universe with the unification of $F (R)$ and $F (T )$ gravity theories as $F(R,T)$ gravity which one is an arbitrary function of curvature scalar $R$ and torsion scalar $T$ \cite{Myrzakulov_2012, Sharif_2012}. The existence of both scalars $R$ and $T$ into $F(R,T)$ gravity is effective on only geometry of Universe no on matter source. This means that Universe is included the effects of curvature and torsion.

As we discussed, the acceleration can be consequence of the dark energy influence that this leads to some other models called Chaplygin gas \cite{Amani_2014, Kamenshchik_2001, Bento_2002, Chimento_2004, Amanib_2013, Naji_2014}. Chaplygin gas is a fluid with negative pressure that begins to dominate the matter content and, at the end, the process of structure formation is driven by cold dark matter without affecting the previous history of the universe. This kind of Chaplygin gas cosmology has an interesting connection to String Theory via the Nambu-Goto action for a D-brane moving in a $(D+2)$-dimensional space-time, feature than can be regarded to the tachyonic panorama \cite{Ogawa_2000}.

The main objective of this paper is that we will consider the $F(R,T)$ gravity model as a source of dark energy. Therefore, we intend to investigate the interacting model between $F(R,T)$ gravity and modified Chaplygin gas with this motivation that  we can describe the accelerated expansion of the Universe.

The paper is organized as follows:

In Sec. \ref{s2}, we review $F(R,T)$ gravity model and obtain the Friedmann equations by using the corresponding connections. In Sec. \ref{s3}, we introduce the basic setup of the modified Chaplygin gas, and then will interact $F(R,T)$ gravity with modified Chaplygin gas. Thereinafter, total energy density and total pressure of Universe Will be written in the form of a combination of modified Chaplygin gas and dark energy. In Sec. \ref{s4}, we reconstruct the current model with a source of scalar field by quintessence model, and then the cosmological parameters will be written in terms of redshift and $e$-folding number. Also we investigate stability of the model and present corresponding graphs in terms of redshift and $e$-folding number. Finally, a short summary is given in Sec. \ref{s5}.

\section{Fundamental of $F(R,T)$ gravity}\label{s2}
The action of $F(R,T)$ gravity theory coupled with matter is given by \cite{Myrzakulov_2012, Jamil_2012, Pasqua_2013}
\begin{equation}\label{S}
S = \int d^{4}x \,e \left(F(R,T) +\mathcal{L}_m \right),
\end{equation}
where $e = det (e^i_{\,\,\mu}) = \sqrt{-g}$, in which $g$ is the determinant of the metric tensor $g_{\mu \nu}$, and $\mathcal{L}_m$ is the matter Lagrangian. The $F(R,T)$ is an arbitrary function of curvature scalar $R$ and torsion scalar $T$. We note that curvature scalar represents gravity in general relativity, and torsion scalar represents gravity in teleparallel gravity by a different mathematical notations as Levi-Civita connection and  Wienzb\"{o}ck connection, respectively. Despite this difference, independently both theories have similar result for equivalent descriptions of gravitation. Therefore, we consider a vierbein field $e_i(x^\mu)$ with index $i$ running from $0$ to $3$, which one is an orthonormal basis for the tangent space at each point $x^\mu$ of the manifold in  Wienzb\"{o}ck connection. Then, we can relate vierbein field to the metric as $g_{\mu \nu} = \eta_{i j} e^i_{\,\,\mu} e^j_{\,\,\nu}$ in which the Minkowski metric $ \eta_{i j}=diag(-1,+1,+1,+1)$.

We consider homogeneous and isotropic Universe in the flat FLRW metric in the following form
\begin{equation}\label{frw}
  ds^2=-dt^2+a^2(t)(dx^2+dy^2+dz^2),
\end{equation}
where $a(t)$ is scale factor, and one is the function of cosmic time. We can write the vierbein field in accordance with \eqref{frw} as follows
\begin{equation}\label{eimu}
  e^i_{\,\,\mu} = diag (1, a, a, a), \,\,\,\,\,\,\,\, e^\mu _{\,\,\,\,i} = diag (1, a, a, a),
\end{equation}
Now we intend to find the corresponding Lagrangian of the action \eqref{S} as
\begin{equation}\label{S1}
  S=\int \mathcal{L} dt,
\end{equation}
in that case, we expand expression $F(R,T)$ by Maclaurin series, so that the Lagrangian is simplicity yielded as
\begin{equation}\label{lag1}
  \mathcal{L}=a^3 \left(F-TF_T-RF_R+vF_T+uF_R\right) - 6\left(F_R+F_T\right) a \dot{a}^2-6\left(\dot{R} F_{RR}+\dot{T} F_{RT}\right)+a^3 \mathcal{L}_m,
\end{equation}
where indices denote derivative with respect to $R$ and $T$ in corresponding locations, and the point denotes derivative with respect to cosmic time. The curvature and torsion scalars are find by
\begin{subequations}\label{RT1}
\begin{eqnarray}
 R &=& u+g^{\mu \nu} R_{\mu \nu}=u+6\left(\dot{H}+2H^2\right), \label{RT1-1}\\
 T &=& v-S_\rho^{\,\,\,\,\mu \nu} T^{\rho}_{\,\,\,\,\mu \nu}=v-6H^2,\label{RT1-2}
\end{eqnarray}
\end{subequations}
where $H=\frac{\dot{a}}{a}$ is the Hubble parameter, and, $u=u(a,\dot{a})$ and $v=v(a,\dot{a})$ be defined as the two arbitrary functions in terms of $a$ and $\dot{a}$, and,  $R_{\mu \nu}$, $S_\rho^{\,\,\,\,\mu \nu}$ and $T^{\rho}_{\,\,\,\,\mu \nu}$ are the Ricci, antisymmetric and torsion tensors respectively, in the form
 \begin{subequations}\label{RST}
\begin{eqnarray}
R_{\mu\nu}&=&\partial_{\lambda}\Gamma^{\lambda}_{\mu\nu}-\partial_{\mu}\Gamma^{\lambda}_{\lambda
\nu}+\Gamma^{\lambda}_{\mu\nu}\Gamma^{\rho}_{\rho\lambda}-\Gamma^{\lambda}_{\nu\rho}
\Gamma^{\rho}_{\mu\lambda},\label{RST-1}\\
S_\rho{}^{\mu\nu}&=&\frac{1}{2}(K^{\mu\nu}{}_\rho+\delta^\mu_\rho
T^{\alpha\nu}{}_\alpha-\delta^\nu_\rho T^{\alpha\mu}{}_\alpha),\label{RST-2}\\
T^{\lambda}~_{\mu\nu}&=&\Gamma^{\lambda}~_{\nu\mu}-
\Gamma^{\lambda}~_{\mu\nu}=e_A^\rho\,(\partial_\mu e^A_\nu-\partial_\nu
e^A_\mu)\label{RST-3},
\end{eqnarray}
\end{subequations}
where the contortion tensor $K^{\mu\nu}{}_\rho$ is
\begin{equation}\label{k1}
K^{\mu\nu}{}_\rho=-\frac{1}{2}(T^{\mu\nu}{}_\rho-T^{\nu\mu}{}_\rho
-T_\rho{}^{\mu\nu}).
\end{equation}
By using the Eq. \eqref{lag1}, we can obtain the Friedmann equations in the following form
\begin{subequations}\label{rhop}
\begin{eqnarray}
&2a^3\rho_{tot} = 6a^2\dot{a}\dot{R}F_{RR}-\left(6a^2\ddot{a}+a^3\dot{a}\frac{\partial u}{\partial \dot{a}}\right)F_R+6a^{2}\dot{a}\dot{T}F_{RT}+\left(12a{\dot{a}}^{2} -a^{3}\dot{a}\frac{\partial v}{\partial \dot{a}}\right)F_T+a^3F,\label{rhop-1}\\
&6a^{2}p_{tot} = -6a^{2}{\dot{R}}^{2}F_{RRR}-\left(12a\dot{a}\dot{R}+6a^{2}\ddot{R}-a^{3}\dot{R}\frac{\partial
u}{\partial \dot{a}}\right)F_{RR}+\Big(12{\dot{a}}^{2}+6a\ddot{a} +3a^{2}\dot{a}\frac{\partial u}{\partial \dot{a}}\label{rhop-2}\\\nonumber
&+a^{3}\frac{\partial}{\partial t}\left(\frac{\partial u}{\partial \dot{a}}\right)-a^{3}\frac{\partial u}{\partial
a}\Big)F_{R}-\left(12a\dot{a}\dot{T}-a^{3}\dot{T}\frac{\partial v}{\partial \dot{a}}\right)F_{TT}
-\Big(24{\dot{a}}^{2}+12a\ddot{a}-3a^{2}\dot{a}\frac{\partial v}{\partial \dot{a}}-a^{3}\frac{\partial}{\partial t}(\frac{\partial v}{\partial \dot{a}})\\\nonumber&+a^{3}\frac{\partial v}{\partial a}\Big)F_{T} -\left(12a\dot{a}\dot{T}+12a\dot{a}\dot{R}+6a^{2}\ddot{T}-a^{3}\dot{R}\frac{\partial v}{\partial \dot{a}}-a^{3}\dot{T}\frac{\partial u}{\partial \dot{a}}\right)F_{RT}-12a^{2}\dot{R}\dot{T}F_{RRT}\\\nonumber
&-6a^{2}{\dot{T}}^{2}F_{RTT}-3a^{2}F,
\end{eqnarray}
\end{subequations}
where $\rho_{tot}$ and $p_{tot}$ are total energy density and total pressure of an Universe dominated with a perfect fluid, respectively. The continuity equation of the model becomes
\begin{equation}\label{cont}
  \dot{\rho}_{tot}+3 H (\rho_{tot}+p_{tot})=0.
\end{equation}
In this paper, we consider a simple particular model of $F(R,T)$ gravity as
\begin{equation}\label{frt}
F(R,T)=\mu R+\nu T,
\end{equation}
where $\mu$ and $\nu$ are constants.

Now for solving the model, we choose a power form for functions $u$ and $v$ as
\begin{subequations}\label{uv1}
\begin{eqnarray}
u &=& \alpha \, a^n,\label{uv1-1} \\
v &=& \beta \, a^m,\label{uv1-2}
\end{eqnarray}
\end{subequations}
to substitute Eqs. \eqref{uv1} into Eqs. \eqref{rhop}, the Friedmann equations are rewritten by
\begin{subequations}\label{rho2}
\begin{eqnarray}
&\rho_{tot}=3(\mu+\nu)H^2+\frac{1}{2}\left(\mu\,\alpha\,a^n+\nu \,\beta\, a^m\right),\label{rho2-1}\\
&-p_{tot}=(\mu+\nu)\left(2\dot{H}+3H^2\right)+\frac{1}{6} \mu\,\alpha(n+3)\, a^n+\frac{1}{6} \nu\,\beta (m+3)\, a^m.\label{rho2-2}
\end{eqnarray}
\end{subequations}

\section{Interacting $F(R,T)$ gravity with modified Chaplygin gas}\label{s3}
One of the recent cosmological models which is based on the use of exotic type of perfect fluid suggests that our Universe filled with the Chaplygin gas with the following equation of state
\begin{equation}\label{pcg}
p_{CG}=-\frac{B}{\rho_{CG}},
\end{equation}
where $B$ is a positive constant.  The equation of state \eqref{pcg} generalized as
\begin{equation}\label{pgcg}
p_{GCG}=-\frac{B}{\rho_{GCG}^{\gamma}},
\end{equation}
which is known as generalized Chaplygin gas.
Therefore, modified Chaplygin gas with the following equation of state introduced
\begin{equation}\label{pmcg}
p_{MCG}=A \rho_{MCG}-\frac{B}{\rho_{MCG}^{\gamma}},
\end{equation}
where $A$ is a positive constant, and $p_{MCG}$ and $\rho_{MCG}$ are the pressure and energy density in which $0 < \gamma < 1$. This model is more appropriate choice to have constant negative pressure at low energy density and high pressure at high energy density. The special case of $A = 1/3$ is the best fitted value to describe evolution of the Universe from radiation regime to the $\Lambda C D M$ regime. Also, the modified Chaplygin gas has been investigated in Refs. \cite{Naji_2014, Debnath_2004, Lu_2008}.

Therefore, we consider the total energy density and pressure of Universe as the combination of components of dark energy and Chaplygin gas in the following form
\begin{subequations}\label{rhoptot}
\begin{eqnarray}
&\rho_{tot}=\rho_{MCG}+\rho_{DE},\label{rhoptot-1}\\
&p_{tot}=p_{MCG}+p_{DE}.\label{rhoptot-2}
\end{eqnarray}
\end{subequations}
Now, we take an energy flow between the dark energy and modified Chaplygin gas, so the energy flow is introduced as an interaction between two components. In this way, the continuity equation \eqref{cont} be written to two separate continuity equations in the following form
\begin{subequations}\label{rhopQ}
\begin{eqnarray}
&\dot{\rho}_{DE}+3H(p_{DE}+\rho_{DE})= -Q,\label{rhopQ-1}\\
&\dot{\rho}_{MCG}+3H(p_{MCG}+\rho_{MCG})= Q,\label{rhopQ-2}
\end{eqnarray}
\end{subequations}
where $Q$ is the interaction term between the dark energy and modified Chaplygin gas, that one be usually considered as: $Q = 3 H b^2 \rho_{DE}$, $Q = 3 H b^2 \rho_{MCG}$ and $Q = 3 H b^2 \rho_{tot}$ in which $b^2$ is the coupling parameter or transfer strength \cite{Chimento_2003, Khurshudyan_2014, Naji_2014, Amani_2013, Amani_2014}. This choice is based on positive motivation $Q$, because from the observational data at the four years $WMAP$ implies that the coupling parameter must be a small positive value \cite{Spergel_2003, Feng_2008}.

In this paper, we consider $Q = 3 H b^2 \rho_{MCG}$, and by substituting $Q$ into Eq. \eqref{rhopQ-2} we can obtain energy density of modified Chaplygin gas as
\begin{equation}\label{rhogcg1}
\rho_{MCG}=\left[\frac{B}{\eta}+c_0 a^{-3 \eta (1+\gamma)}\right]^{\frac{1}{1+\gamma}},
\end{equation}
where $\eta = 1+A-b^2$, and then
\begin{eqnarray}\label{pgcg1}
p_{MCG} = A \left[\frac{B}{\eta}+c_0 a^{-3 \eta (1+\gamma)}\right]^{\frac{1}{1+\gamma}}-\frac{B}{\left[\frac{B}{\eta}+c_0 a^{-3 \eta (1+\gamma)}\right]^{\frac{\gamma}{1+\gamma}}},
\end{eqnarray}
where $c_0$ is a constant of integration.

The $\rho_{DE}$ and $p_{DE}$ obtain by Eqs. \eqref{rho2} and \eqref{rhoptot} in the following form
\begin{subequations}\label{prhode}
\begin{eqnarray}
&\rho_{DE}=3(\mu+\nu)H^2+\frac{1}{2}\left(\mu\,\alpha\,a^n+\nu \,\beta\, a^m\right) -\left[\frac{B}{\eta}+c_0 a^{-3 \eta (1+\gamma)}\right]^{\frac{1}{1+\gamma}},\label{prhode-1}\\
&p_{DE}=-(\mu+\nu)\left(2\dot{H}+3H^2\right)-\frac{1}{6} \mu\,\alpha(n+3)\, a^n-\frac{1}{6} \nu\,\beta (m+3)\, a^m \label{prhode-2}\\ \nonumber
 &-A \left[\frac{B}{\eta}+c_0 a^{-3 \eta (1+\gamma)}\right]^{\frac{1}{1+\gamma}}+
\frac{B}{\left[\frac{B}{\eta}+c_0 a^{-3 \eta (1+\gamma)}\right]^{\frac{\gamma}{1+\gamma}}}.
\end{eqnarray}
\end{subequations}
The dark energy EoS becomes
\begin{equation}\label{omegade}
\omega_{DE}=\frac{p_{DE}}{\rho_{DE}}.
\end{equation}

\section{Reconstructing and parametrization}\label{s4}
\begin{figure}[t]
\begin{center}
\includegraphics[scale=.3]{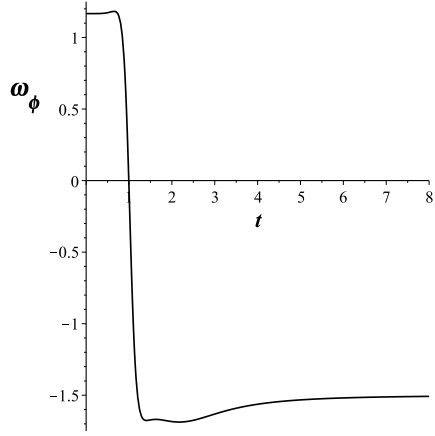}
\caption{The graphs of $\omega_\phi$ in terms of cosmic time for $A = 0.5$, $B = 0.5$, $c_0 = 0.5$, $b = 0.5$, $\gamma = 0.5$, $a_0 = 1$, $s = 2$, $\alpha = 1$, $\beta = 1$, $\mu = 0.5$, $\nu = 0.5$, $n = 1.5$ and $m = -6.5$.}\label{omegaphi2t}
\end{center}
\end{figure}

In this section, we first will reconstruct the model as a scalar field and a potential entitled the quintessence model. For this purpose, we suppose that the origin of interacted $F(R,T)$ gravity with Chaplygin gas is a real scalar field in which the corresponding Friedmann equations is written in the following form \cite{Copeland_2006, Dent_2011}
\begin{subequations}\label{rhopphi}
\begin{eqnarray}
 &\rho_{\phi}=\frac{3}{\kappa^2}H^2=\frac{1}{2}\dot{\phi}^2+V(\phi),\label{rhopphi-1}\\
 &p_{\phi}=-\frac{1}{\kappa^2}(3H^2+2\dot{H})=\frac{1}{2}\dot{\phi}^2-V(\phi),\label{rhopphi-2}
\end{eqnarray}
\end{subequations}
the continuity equation and the EoS parameter are given by
\begin{subequations}\label{rhoeos}
\begin{eqnarray}
&\dot{ \rho}_{\phi}+3H(1+\omega_{\phi})\rho_{\phi}=0,\label{rhoeos-1}\\
&\omega_{\phi}=-1-\frac{2}{3}\frac{\dot{H}}{H^2}=\frac{\dot{\phi}^2-2V(\phi)}{\dot{\phi}^2+2V(\phi)}.\label{rhoeos-2}
\end{eqnarray}
\end{subequations}
Now let's replace Eqs. \eqref{rhoeos} by transforming $\rho_{\phi}\rightarrow \rho_{DE}$ and $p_{\phi}\rightarrow p_{DE}$ into \eqref{prhode}, so that we find kinetic energy term and potential as
\begin{subequations}\label{phidotpphi}
\begin{eqnarray}
&\dot{\phi}^2 = \rho_{\phi} + p_{\phi} = 2 (\mu+\nu) \dot{H}-\frac{1}{6} \mu \alpha n a^n -\frac{1}{6} \nu \beta m a^m \label{phidotpphi-1}\\ \nonumber
 &-\left[\frac{B}{\eta}+c_0 a^{-3 \eta (1+\gamma)}\right]^{\frac{1}{1+\gamma}}\left(1+A-\frac{B}{\frac{B}{\eta}+c_0 a^{-3 \eta (1+\gamma)}}\right),\\
&V(\phi) = \frac{1}{2}(\rho_{\phi} -p_{\phi}) = (\mu+\nu) \left(\dot{H}+3 H^2\right)+\frac{1}{12} \mu \alpha (n+6) a^n +\frac{1}{12} \nu \beta (m+6) a^m \label{phidotpphi-2}\\ \nonumber
 &-\frac{1}{2}\left[\frac{B}{\eta}+c_0 a^{-3 \eta (1+\gamma)}\right]^{\frac{1}{1+\gamma}}\left(1-A+\frac{B}{\frac{B}{\eta}+c_0 a^{-3 \eta (1+\gamma)}}\right),
\end{eqnarray}
\end{subequations}
the right hand side of Eqs. \eqref{phidotpphi} are determined as a function of the scale factor. In that case, we take a particular form of Universe evolution by power law for the scale factor in terms of cosmic time as follows:
\begin{equation}\label{a(t)1}
  a(t) = a_0 t^s,
\end{equation}
where $a_0$ is the value of the scale factor at the present Universe, and $s>1$ is consistent the accelerated expansion of Universe. To substitute the aforesaid scale factor into Eqs. \eqref{phidotpphi}, we can obtain the following relationships
\begin{subequations}\label{phidotpphi1}
\begin{eqnarray}
&\dot{\phi}^2 = -2 (\mu+\nu) \frac{s}{t^2}-\frac{1}{6} \mu \,\alpha\,n \,a_0^n\, t^{ns} -\frac{1}{6} \nu\, \beta\,m\, a_0^m\, t^{ms} \label{phidotpphi1-1}\\ \nonumber
 &-\left[\frac{B}{\eta}+c_0 (a_0\, t^s)^{-3 \eta (1+\gamma)}\right]^{\frac{1}{1+\gamma}}\left(1+A-\frac{B}{\frac{B}{\eta}+c_0 (a_0\, t^s)^{-3 \eta (1+\gamma)}}\right),\\
&V(\phi) = (\mu+\nu) \frac{s(3s-1)}{t^2}+\frac{1}{12} \mu \,\alpha\, (n+6)\, a_0^n\,t^{ns} +\frac{1}{12} \nu\, \beta\, (m+6)\, a_0^m\,t^{ms} \label{phidotpphi1-2}\\ \nonumber
 &-\frac{1}{2}\left[\frac{B}{\eta}+c_0 (a_0\, t^s)^{-3 \eta (1+\gamma)}\right]^{\frac{1}{1+\gamma}}\left(1-A+\frac{B}{\frac{B}{\eta}+c_0 (a_0\, t^s)^{-3 \eta (1+\gamma)}}\right).
\end{eqnarray}
\end{subequations}
The Eq. \eqref{phidotpphi1-2} is a reconstructed potential. This means that Universe is described by the quintessence model with the field $\phi(t)$ and the potential $V(\phi)$ instead of $F(R,T)$ gravity. Here we realize that interacting $F(R,T)$ gravity with modified Chaplygin gas is a good alternative to quintessence model. Also this reconstruction demonstrates the reconstructed EoS of scaler field in Fig. \ref{omegaphi2t}. We note that choice of free parameters play a very important role in describing this model, so the motivation for these choices is based on crossing over phantom divide line.

In what follows, we turn our attention to study our model by parametrization. In the sense that the energy density and pressure of dark energy are written in terms of redshift,  $z=\frac{a_0}{a(t)}-1$, in which $a_0$ is the value of the scale factor at the present Universe, in other words, $z=0$ corresponds the today Universe. We introduce parameter, $r=\frac{H^2}{H_0^2}$ in which $H_0=71\pm3\,km\,s^{-1}\,Mpc^{-1}$ is the value of Hubble parameter of current Universe. In that case, we find
\begin{equation}\label{ddt}
  \frac{d}{dt}=-(1+z) H \frac{d}{dz},
\end{equation}
then, we can rewrite the Eqs. \eqref{prhode} and \eqref{omegade} in terms of $z$ as
  \begin{subequations}\label{prhode1}
\begin{eqnarray}
&\rho_{DE}=3 H_0^2 (\mu+\nu) r+\frac{1}{2}\left(\mu\,\alpha\,\left(\frac{a_0}{1+z}\right)^n+\nu \,\beta\, \left(\frac{a_0}{1+z}\right)^m\right)\label{prhode1-1}\\ \nonumber &-\left[\frac{B}{\eta}+c_0 \left(\frac{a_0}{1+z}\right)^{-3 \eta (1+\gamma)}\right]^{\frac{1}{1+\gamma}},\\
&p_{DE}=-H_0^2 (\mu+\nu)\left(-(1+z) r'+3 r\right)-\frac{1}{6} \mu\,\alpha(n+3)\, \left(\frac{a_0}{1+z}\right)^n-\frac{1}{6} \nu\,\beta (m+3)\, \left(\frac{a_0}{1+z}\right)^m \label{prhode1-2}\\ \nonumber
 &-A \left[\frac{B}{\eta}+c_0 \left(\frac{a_0}{1+z}\right)^{-3 \eta (1+\gamma)}\right]^{\frac{1}{1+\gamma}}+
\frac{B}{\left[\frac{B}{\eta}+c_0 \left(\frac{a_0}{1+z}\right)^{-3 \eta (1+\gamma)}\right]^{\frac{\gamma}{1+\gamma}}},\\
&\omega_{DE}=-1+ \label{prhode1-3}\\ \nonumber
&\frac{H_0^2 (\mu+\nu) (1+z) r'-\frac{1}{6} \mu \alpha n \left(\frac{a_0}{1+z}\right)^n-\frac{1}{6} \nu \beta m \left(\frac{a_0}{1+z}\right)^m-(1+A) \left[\frac{B}{\eta}+c_0 \left(\frac{a_0}{1+z}\right)^{-3 \eta (1+\gamma)}\right]^{\frac{1}{1+\gamma}}+\frac{B}{\left[\frac{B}{\eta}+c_0 \left(\frac{a_0}{1+z}\right)^{-3 \eta (1+\gamma)}\right]^{\frac{\gamma}{1+\gamma}}}}{3 H_0^2 (\mu+\nu) r+\frac{1}{2} \mu \alpha \left(\frac{a_0}{1+z}\right)^n+\frac{1}{2} \nu \beta \left(\frac{a_0}{1+z}\right)^m-\left[\frac{B}{\eta}+c_0 \left(\frac{a_0}{1+z}\right)^{-3 \eta (1+\gamma)}\right]^{\frac{1}{1+\gamma}}},
\end{eqnarray}
\end{subequations}
where the prime denotes a derivative with respect to $z$.

\begin{figure}[t]
\begin{center}
\subfigure
{\includegraphics[scale=.3]{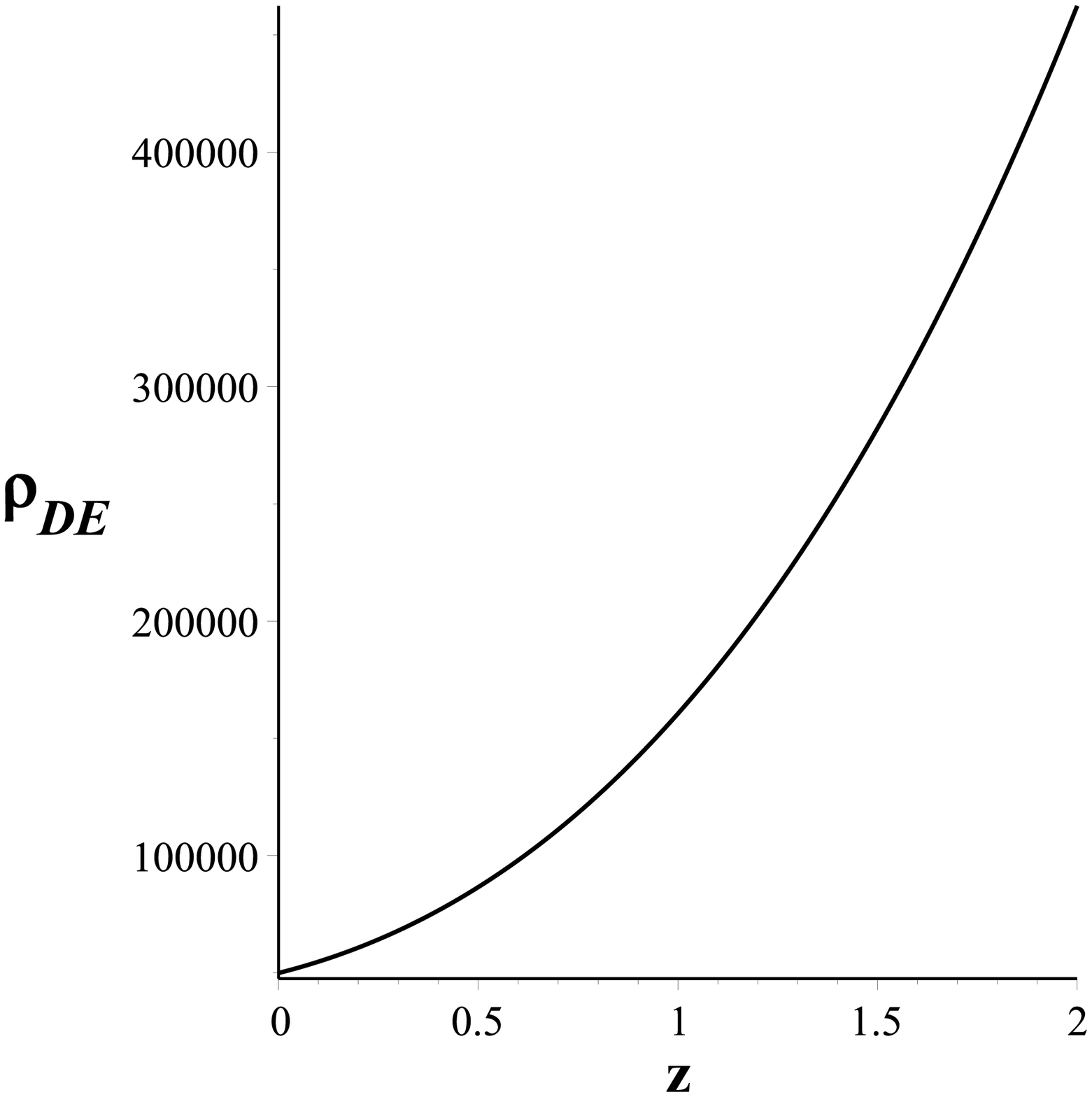}\label{rhiDE}}
\subfigure
{\includegraphics[scale=.3]{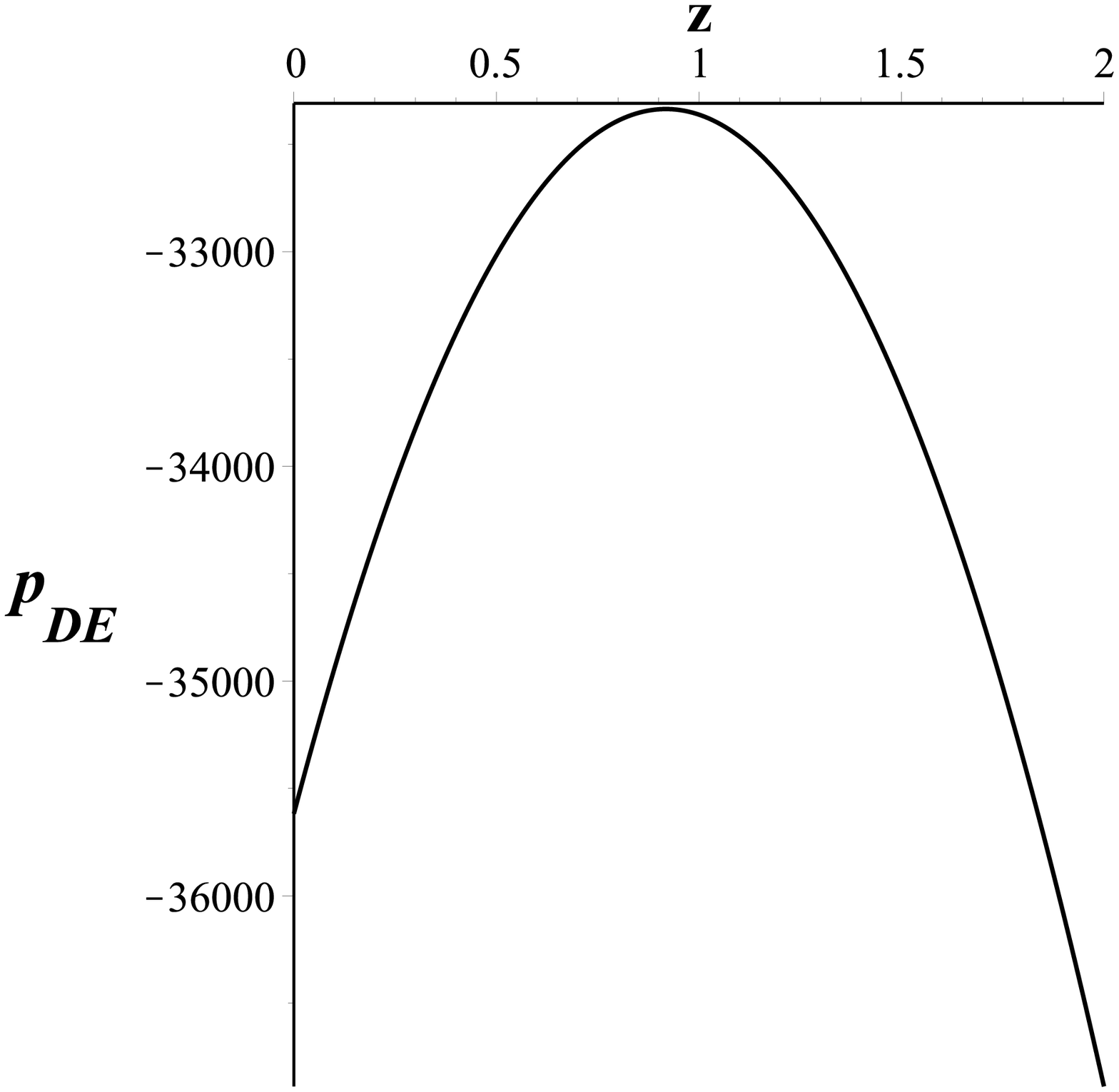}\label{pDE}}
\caption{The graphs of $\rho_{DE}$ and $p_{DE}$ in terms of redshift $z$ for $A = 0.5$, $B = 5$, $c_0 =  1$, $b = 0.95$, $\gamma = 0.25$, $a_0 = 0.55$, $\alpha = 3$, $\beta = 2$, $\mu = 1$, $\nu = 2$, $n = 2$ and $m = -2$.}\label{rhopDE}
\end{center}
\end{figure}
Now, in order to describe the model as a origin of dark energy, we intend to reconstruct the EoS of the model by parametrizing the Hubble parameter in terms of the
 $z$. We note that reconstruction of dark energy models was fulfilled in Refs. \cite{Starobinsky_1998, Huterer_1999, Alam_2007, Sadeghi_2009, Setare_2009} for a minimally scalar field with a potential. Therefore, we take the following parametrization as
\begin{equation}\label{r(z)}
r(z) = \Omega_{0m} (1+z)^3+A_0 +A_1 (1+z) +A_2 (1+z)^2,
\end{equation}
where values of the corresponding coefficients are $\Omega_{0m}=0.30$, $A_0=1$, $A_1=-0.48$ and $A_2=0.25$, which these obtained the results of Refs. \cite{Wang_2006, Gong_2007}.

\begin{figure}[t]
\begin{center}
\includegraphics[scale=.35]{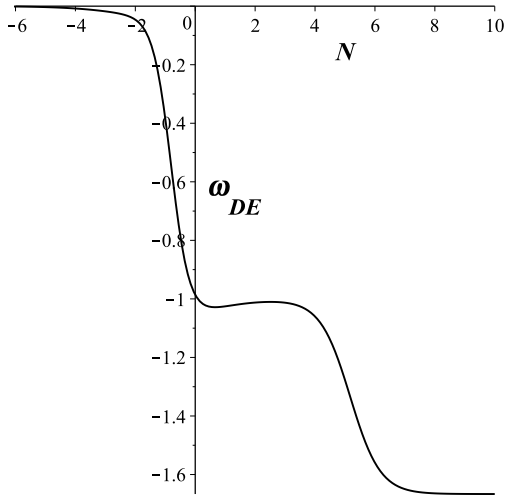}
\caption{The graph of $\omega_{DE}$ in terms of $e$-folding number $N$ for $A = 0.5$, $B = 5$, $c_0 =  1$, $b = 0.95$, $\gamma = 0.25$, $a_0 = 0.55$, $\alpha = 3$, $\beta = 2$, $\mu = 1$, $\nu = 2$, $n = 2$ and $m = -2$.}\label{omegaDE}
\end{center}
\end{figure}

By inserting Eq. \eqref{r(z)} into Eqs. \eqref{prhode1}, expressions $\rho_{DE}$, $p_{DE}$ and $\omega_{DE}$ will obtain in terms of redshift $z$. Then, we plot the cosmological parameters in Figs. \ref{rhopDE} and \ref{omegaDE} by the natural units $c = \hbar = 16 \pi G = 1$. We note that free parameters play the role of an important to plot the cosmological parameters, in which the motivation of the selections is based on positivity energy density and negativity pressure.

Now for that free parameters correspond with observable quantities, we plot the EoS of the model versus the $e$-folding number, N = ln(a) as the time variable in Fig. \ref{omegaDE}. We can see the values of EoS in three cases $N \rightarrow -\infty$, $N = 0$ ( late time) and $N \rightarrow +\infty$ respectively with values $0$, $-0.98$ and $-1.67$. Therefore, Fig. \ref{omegaDE} shows us that variation of the EoS from early time till future time is $0$ till $-1.67$, and this confirms the existence of an accelerated expansion of the Universe. The value of EoS of late time confirms the result of \cite{Amanullah_2010} in a flat Universe.

In order to have a more complete discussion, we intend to discuss the stability of the model. For this purpose, we will use an useful function $c^2_s=\frac{\partial_z p_{DE}}{\partial_z \rho_{DE}}$ (in which $\partial_z=\frac{\partial}{\partial z}$) as a thermodynamic system, so this function represents sound speed in a perfect fluid. Therefore, the stability condition occurs when the function $c^2_s$ becomes bigger than zero. It should be noted that a thermodynamic system can be described with adiabatic and non-adiabatic perturbations. This means that the aforesaid function $c^2_s $ be related to an adiabatic sound speed.

\begin{figure}[t]
\begin{center}
\includegraphics[scale=.35]{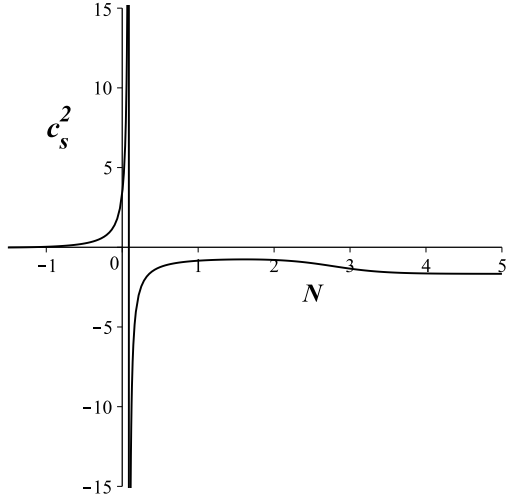}
\caption{The graph of $c_s^2$ in terms of $e$-folding number $N$ for $A = 0.5$, $B = 5$, $c_0 =  1$, $b = 0.95$, $\gamma = 0.25$, $a_0 = 0.55$, $\alpha = 3$, $\beta = 2$, $\mu = 1$, $\nu = 2$, $n = 2$ and $m = -2$.}\label{c2s2N}
\end{center}
\end{figure}

For making derivative the Eqs. \eqref{prhode1-1} and \eqref{prhode1-2} with respect to redshift $z$, the speed sound function be plotted in terms of $e$-folding number $N$ in Fig. \ref{c2s2N}. We can see in the corresponding figure the values of $c^2_s$ in three cases $N \rightarrow -\infty$, late time ($N = 0$) and $N \rightarrow +\infty$ respectively with values $0$, $3.4$ and $-1.67$. Therefore, the Fig. \ref{c2s2N} shows us that there is a stability in late time, because value $c^2_s$ is positive for case $N=0$.

\section{conclusion}\label{s5}
In this paper, we have studied $F(R,T)$ gravity in terms of an arbitrary function of curvature and torsion scalars. For simplicity, the function $F(R,T)$ has been gotten as a linear function of the curvature and torsion scalars, and have computed in FLRW metric. The motivation of this research is based on existence of  both cases curvature and torsion in the Universe, while so far the research carried out on just implications existence of the curvature in the Universe. Afterwards, we have considered the $F(R,T)$ gravity model by interacting with modified Chaplygin gas. We wrote the corresponding action as the combination of $F(R,T)$ gravity and a matter Lagrangian. In order to obtain the friedmann equations, we used a different mathematical notations for curvature scalar and torsion scalar by Levi-Civita connection and Wienzb\"{o}ck connection, respectively.

In what follows, we considered total energy density and total pressure of Universe as dominated with a perfect fluid, in which dark energy contribution is taken the combination of $F(R,T)$ gravity and modified Chaplygin gas as Eq. \eqref{rhoptot}. Then, by dividing these two functions we obtained the EoS of dark energy for the scenario.

Subsequently, we reconstructed the model by an origin of a scalar field entitled quintessence model with a field potential. With correspondence between our model and quintessence model, we plotted the EoS of scalar field in terms of cosmic time by taking the scale factor as power law. The graph of corresponding EoS described the accelerated expanding of Universe. Next, we used a parametrization \eqref{r(z)}, and the cosmological parameters have been written in terms of redshift $z$. In that case, we drew them with respect to $z$ and $e$-folding number $N$ in Figs. \ref{rhopDE} and \ref{omegaDE}. These figures showed us the existence of an accelerating Universe, especially graph of EoS showed us that its value in late time is $-0.98$ which one is consistent with Ref. \cite{Amanullah_2010}. We noted that the free parameters play an important role in plotting figures, the motivation of these selections are based on positivity energy density and negativity pressure.  Finally, in order to investigate the stability, we tried to compute speed sound in this scenario in terms of $e$-folding number, so we plotted the $c_s^2$ with respect to $e$-folding number and have obtained its corresponding values in three status $N \rightarrow -\infty$, late time ($N = 0$) and $N \rightarrow +\infty$. Therefore, the Fig. \ref{c2s2N} showed us that there is the stability in late time.


\end{document}